\documentclass{iaus}
\usepackage{graphicx}

\title[Dust emission from PN IC418] 
{Modeling the dust emission from PN IC418}

\author[C. Morisset, R. Szczerba, D. A. Garc\'{\i}a-Hern\'andez \& P. Garc\'{\i}a-Lario]   
{Christophe Morisset$^{1,2}$,
 Ryszard Szczerba$^3$,
 D. A. Garc\'{\i}a-Hern\'andez$^{2}$,
 \and Pedro Garc\'{\i}a-Lario$^4$}

\affiliation{
$^1$Instituto de Astronom\'{\i}a, Universidad Nacional Autonoma de M\'exico.  {\tt Chris.Morisset@Gmail.com} \\[\affilskip]
$^2$Instituto de Astrof\'{\i}sica de Canarias (and ULL), E-38200, La Laguna, Tenerife, Spain. \\[\affilskip]
$^3$Nicolaus Copernicus Astronomical Center, Toru\'n, Poland \\[\affilskip]
$^4$Herschel Science Centre, European Space Astronomy Centre, Madrid, Spain} 

\pubyear{2011}
\volume{283}  
\pagerange{}
\jname{Planetary Nebulae: An Eye to the Future}
\editors{A. Manchado, \& L. Stanghellini, eds.}
\begin{document}

\maketitle

\begin{abstract}
We construct a detailed model for the IR dust emission from the PN IC 418. We succeed to reproduce the emission from 2 to 200$\mu$m. We can determine the amount of emitting dust as well as its composition, and compare to the depletion of elements determined for the photoionized region.
\end{abstract}

\firstsection 
\section{Introduction}
The depletion of some elements in the gaseous phase of the Planetary Nebulae (PNe) is attributed to their trapping in dust grains. In this work we compare the abundances obtained in the gaseous phase and in the grains for the PN IC418.
\cite{2009Morisset_aap507} performed a detailed model of this PN, combining self-consistent photoionization and atmosphere models to reproduce all the available observational material (Stellar continuum and lines, nebular images and lines intensities).
The main properties of the model (performed using the Cloudy code from \cite{1998Ferland_pasp110}) are taken from MG09 and are the following: the nebula has a double shell morphology with a high density PDR almost in pressure equilibrium with the H$^+$ region to reproduce the [CII]157$\mu$m and [OI]63, 145$\mu$m lines. The stellar SED is obtain from a 37.4~kK, 5400 solar luminosity CMFGEN \cite{1998Hillier_apj496} model. See MG09 for more details.

\begin{figure*}
\begin{center}
 \includegraphics[width=12cm]{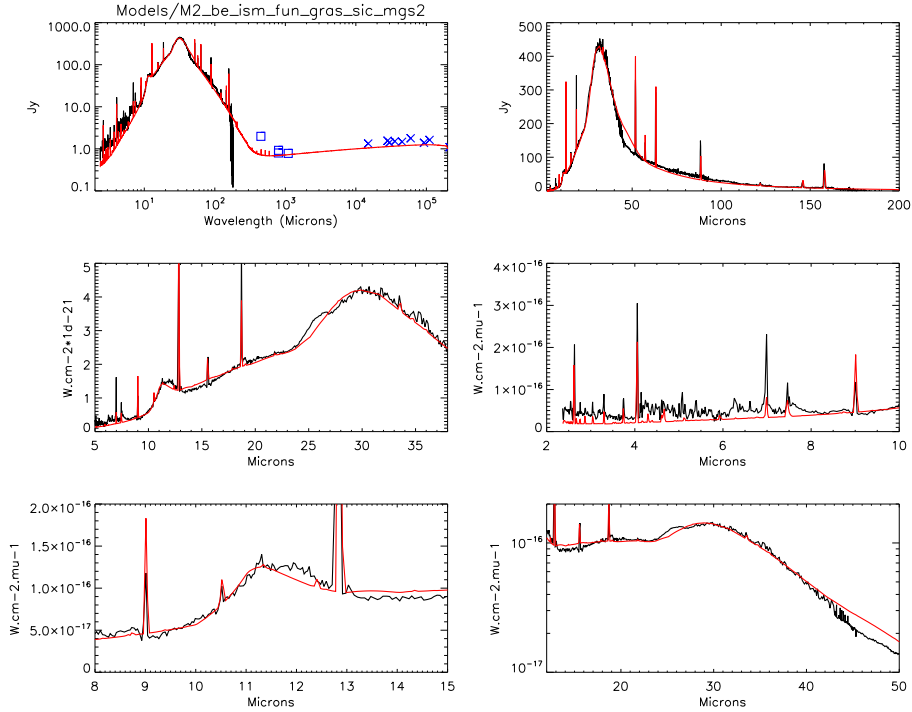} 
 \caption{Comparison between observations (Black: ISO data, Blue: submm and Radio data) and the model (Red). The different panels show the fit in different wavelength ranges and using different units.}
   \label{fig1}
\end{center}
\end{figure*}

\section{Dust emission model}
The global IR emission between 8 and 300$\mu$m is mainly due to amorphous carbon of BE1 type \cite{1991Rouleau_apj377}. We first compute a model where the size distribution of the grains is following a -3.5 power law, with sizes ranging from 0.05 to 0.250$\mu$m. If the dust if present in the whole (H$^+$ and H$^0$) nebula, the resulting emission is too hot.
To reduce the temperature of the grains, we test two options: 1) increasing the mean size of the grains, ranging from 0.06 to 0.400$\mu$m, or 2) putting the dust only in the H$^0$ region. Both hypothesis succeed in shifting the emission to longer wavelengths.

To reproduce the NIR emission, a small grains (0.0005 to 0.01$\mu$m graphitic grains) component is needed.

Both 11.5 and 30$\mu$m features are clearly observed in the IR spectrum of IC418. We explore the emission from SiC and MgS as usual suspects for these features. 
SiC is obtained by including \cite{1988Pegourie_aap194} dielectric data in Cloudy. We use a size distribution with bigger grains (0.06 to 0.400$\mu$m) to slightly shift the emission to longer wavelengths. The fit is still not perfect and we may try other optical properties when available.
For the 30$\mu$m broad feature, we adopt MgS spherical grains (Mie theory) and Continuous Distribution of Ellipsoids approximation (see \cite{1997Szczerba_aap317}).
The CDE approximation leads to better fits to the observations, but we still don't reproduce the observed sub-structures at 21, 26 and 30$\mu$m (also observed by KAO). We may conclude that MgS is not responsible for these emissions (maybe HACs emission, see \cite{2010Garcia-Hernandez_apjl724}), or that better optical data are needed. C$_{60}$ is marginally detected, but we don't try to fit its features. 
\section{Discussion and Conclusion}
We compare the abundances of the depleted material obtained from the photoionization models for the nebular and for the dusty phase.
None of the total abundances for the depleted elements (Si, Mg, S and Fe) are found to be more than solar (the nebula is close to solar metallicity). More details will be published in a forthcoming paper.

\end{document}